\documentclass{ws-procs975x65}            % default, citations in superscript

\newcommand{\sgn}{\operatorname{sgn}}
\newcommand{\diverg}{\operatorname{div}}

\begin{document}

\title{Cosmic cable}

\author{Colin MacLaurin$^*$}

\address{School of Mathematics and Physics,\\
University of Queensland, Brisbane, Australia\\
$^*$E-mail: colin.maclaurin@uqconnect.edu.au}

\begin{abstract}
I investigate the relativistic mechanics of an extended ``cable'' in an arbitrary static, spherically symmetric spacetime. Such hypothetical bodies have been proposed as tests of energy and thermodynamics: by lowering objects toward a black hole, scooping up Hawking radiation, or mining energy from the expansion of the universe. I review existing work on stationary cables, which demonstrates an interesting ``redshift'' of tension, and extend to a case of rigid motion. By using a partly restrained cable to turn a turbine, the energy harvested is up to the equivalent of the cable's rest mass, concurring with the quasistatic case. Still, the total Killing energy of the system is conserved.
%This scenario is insightful for foundational concepts, and investigates little-studied aspects of extended solid objects.
% cut: Intuitively, the energy comes from loss of gravitational potential
\end{abstract}
%\keywords{Style file; \LaTeX; Proceedings; World Scientific Publishing.}

\bodymatter

\section{Introduction}

Hypothetical ropes/strings/cables/tethers in curved spacetime have been applied as test scenarios for energy, thermodynamics, and mechanics: hence are related to major open research questions. The cable considered here obeys ordinary energy conditions for macroscopic matter, and is conceived as engineered rather than naturally occurring. These properties are distinct from other strings in theoretical physics including Nambu-Goto relativistic strings, cosmic strings, and quantum strings such as in superstring theory.
% --- typically concerning black holes or expanding universes ---
% Other loose antecedents might include exact solutions since the Bach-Weyl gravitating rod, Tolman's pressure gradients, or even Rindler rigid acceleration.
% Nambu strings 1970s
% Bach Weyl 1922
% Tolman 1934, also the Tolman-Oppenheimer-Volkoff equation can be derived similarly -- Wikipedia uses the equivalent of Equation (2), combined with an Einstein field equation
% Tolman considered lots of pressure gradients for static spherically symmetric bodies,\cite{tolman1934} and the
% \cite{zwiebach2009}

Penrose proposed lowering a mass towards a black hole to extract energy.\cite{penrose1969} Bekenstein considered lowering a hot object towards a black hole, to convert heat into work and violate the second law of thermodynamics, inspired by a talk of Geroch.\cite{bekenstein1972} However Gibbons was the first to analyse the cable itself, including its tension.\cite{gibbons1972} Unruh \& Wald suggested a resolution to Geroch's thermodynamic paradox, now in the context of Hawking radiation, and showed a ``redshift'' of tension along the cable.\cite{unruhwald1982} Redmount used Weyl's axisymmetric solution to analyse distortions of Schwarzschild spacetime by two masses suspended on ropes.\cite{redmount1984}
% Hence the cable is peripherally linked to various classic papers.
% Penrose was part of his classic review on gravitational collapse
% Geroch talk was 1971
% (This is not Bekenstein's famous black hole thermodynamics paper, which is from 1973.)
% Redmount, the masses are pointlike and static. This was part of the membrane paradigm research program

Others considered cables in Friedmann-Lema\^itre-Robertson-Walker spacetime, to test the dynamics of an expanding universe. Davies showed the futility of scooping up Hawking radiation from the de Sitter horizon.\cite{davies1984mining} Harrison proposed latching a rope onto distant receding bodies, to mine mechanical energy from the expansion.\cite{harrison1995} Others also considered a ``tethered galaxy'',\cite{davis+2003} or a network of strings reminiscent of Szekeres' gravitational compass. There are plenty of other sources, even omitting the cosmic string and string theory literature.
% Sciama 1976 \S3 e.8. Seems to be basically Bekenstein's proposal
% Maugin looks more like Penrose' intent
% Wiltshire 2009 also mentions his "semi-tethered lattice", but says it is a condensed version of the 2008 paper
% Szekeres gravitational springs, nicely summarised in de Felice Bini 2010 p.135. I had cited here \cite{letelier1981,wiltshire2009}
% tethered galaxy: I cut \cite{peacock2001,gronelgaroy2007
% other sources: \cite{sciama1976,maugin1978,wernig-pichler2006}

\section{Newtonian cables}
\label{sec:newtonian}

To show the concept is perfectly reasonable, consider a spool of cable tied to a tree branch at height $L$ above ground. Set up a steady state where a segment always stretches from winch to ground, and moves downwards at constant speed $\beta$. Then under Galilean gravitational acceleration $g$ and linear mass density $\mu$, a constant force $\mu gL$ is exerted downwards. This gives a power $\mu\beta gL$ which can be harnessed at the winch, at the expense of gravitational potential energy.

Next consider a Newtonian ``de Sitter'' cosmology: an empty universe with cosmological constant $\Lambda$. Suppose the winch is anchored at the coordinate origin. Poisson's law $\vec\nabla^2\Phi + \Lambda = 4\pi G\rho$ has solution $\Phi = -\Lambda(x^2+y^2+z^2)/6$ in this case, where $\rho = 0$ since we ignore the cable's own gravitational field. The force on a static cable particle of mass $m$ at location $(x,y,z)$ is $\vec f = -m\,\vec\nabla\Phi = (x,y,z)m\Lambda/3$. The total force at the winch is hence $-\Lambda\mu L^2/6$ for a cable of length $L$. By allowing the cable to recede, work is gained, but the cable is gradually lost to space.
% p.33 of long ``Cosmic cable'' PDF
% can also just get the force by evaluating the potential at the cable's end $(L,0,0)$ say: $\mu\Phi|_{(L,0,0)}

\section{Stationary cable}

This section reviews results of Gibbons and others, with some additions.\cite{gibbons1972,redmount1984,fouxon+2008} Consider an arbitrary spacetime containing a cable with linear mass density $\mu(\mathbf x)$ and tension $T(\mathbf x)$. With no sideways rigidity, the cable's stress-energy tensor is $\mathbf T = \tilde\mu\mathbf u\otimes\mathbf u + \tilde T\mathbf q\otimes\mathbf q$. Here $\mathbf u$ is the $4$-velocity field of cable particles, $\mathbf q$ is a unit spatial vector field pointing along the cable, and $\tilde\mu := \mu/A$ and $\tilde T := T/A$ are scaled by the cross-sectional area.

Conservation of stress-energy implies the vector $\diverg\mathbf T$ vanishes, which when contracted with $\mathbf u$ and $\mathbf q$ leads to $d\tilde\mu/d\tau = -\tilde\mu\diverg\mathbf u + \tilde T\dot{\mathbf q}\cdot\mathbf u$ and $d\tilde T/dL = -\tilde T\diverg\mathbf q -\tilde\mu\mathbf q\cdot\dot{\mathbf u}$ respectively. Here $L$ is the proper-frame length along the cable, $\dot{\mathbf u} := \nabla_{\mathbf u}\mathbf u$ is $4$-acceleration, $\dot{\mathbf q} := \nabla_{\mathbf q}\mathbf q$, and a dot between vectors implies the metric inner product. However stress-energy need not be conserved, for instance if the cable's gravitational field is ignored so $\mathbf T$ is not the source term in the Einstein field equations. Still, $\mathbf u\cdot\diverg\mathbf T = 0$ is guaranteed.\cite{wernig-pichler2006}
% Wernig-Pichler p.39. \S3.1.4, \S4.1.1 ``The energy-momentum tensor is divergence-free, if the configuration is stress-free and the corresponding four-velocity is geodesic.'' ``The divergence of the energy-momentum tensor (this time for a general configuration) is annihilated by the material velocity.'' p.55 ``A configuration, which solves the Euler-Lagrange equation has a divergence-free energy-momentum tensor and vice versa.''
% Tolman textbook see \S95 p.243-4 e.259-60, as cited in Wikipedia on TOV equation. Derivation of TOV from conservation in radial direction, see Wikipedia: https://en.wikipedia.org/wiki/Tolman%E2%80%93Oppenheimer%E2%80%93Volkoff_equation#Derivation_from_general_relativity

Now suppose the spacetime contains a timelike Killing vector field $\boldsymbol\xi$, and that the cable particles are stationary. Hence $\mathbf u = \boldsymbol\xi/V$, where $V := \sqrt{-\boldsymbol\xi\cdot\boldsymbol\xi}$ is the redshift factor. In this case stress-energy conservation is guaranteed. It follows $d\tilde\mu/d\tau = 0$, and if $\mu$ is also constant over space then the tension varies as
\begin{equation}
\boxed{T = W\frac{V_\textrm{end}}{V} + \mu\Big(1-\frac{V_\textrm{end}}{V}\Big)}
\end{equation}
where $W$ is a weight (mass times magnitude of proper acceleration, if a point mass) hanging at the end. This exhibits a curious effect: the tension is ``redshifted'' by transmission along the cable, due to gravitational time-dilation. This is most evident for a massless cable $\mu \equiv 0$,\cite{unruhwald1982} a limiting case of which is familiar in surface gravity as the ``force at infinity'' to support a unit mass particle at a horizon.\cite{wald1984} The cases $W = \mu$, or a cable ending at the horizon with $W = 0$, both yield a constant tension $T \equiv \mu$.\cite{redmount1984,brown2013} Alternatively, one can extend conservation of energy arguments used by Einstein and Bondi for redshift of photons, to justify redshift of force.
% Wald Problem 4 of chapter 6 (e.167), briefer mention in section 12.5, e.339
% Redmount e.114. \S2.4B. Unruh & Wald \S2
% , which realistically requires $w = 0$,

\section{Moving cable}

This section describes the kinematics and dynamics of a moving cable. For further explanation see $\S7$, $\S9$, and $\S11$ of preliminary work.\cite{maclaurin2015}

\subsection{Apparatus and background spacetime}

Consider a static, spherically symmetric spacetime with general metric
\begin{equation}
ds^2 = -e^{2\alpha(r)}dt^2 + e^{2\zeta(r)}dr^2 +r^2(d\theta^2\sin^2\theta\,d\phi^2).
\end{equation}
Static observers form a useful reference or comparison, these have $4$-velocity
\begin{equation}
u_\textrm{static}^\mu = \big(e^{-\alpha},0,0,0\big),
\end{equation}
where the $r$-dependence is omitted for brevity. The $4$-acceleration is $a^\mu = (0,\alpha'e^{-2\zeta},0,0)$ with magnitude $|\alpha'|e^{-2\zeta}$, and is directed towards increasing $r$ if $\alpha' > 0$ and vice versa. A dash will always indicate derivative with respect to $r$. The gravitational redshift factor, determined as usual by static observers, is $\sqrt{-\partial_t\cdot\partial_t} = e^\alpha$.

Assume the cable is hanging ``downwards'' in an equilibrium state, so as a whole the $4$-velocity field is unchanged by translation in $\partial_t$. As in Section~\ref{sec:newtonian}, set a winch or spool at the ``top'' $r = r_0$ say, attached to a turbine which allows only a fixed angular velocity. At the winch location, we assume $e^{\alpha(r_0)} = 1$, which interprets the winch as free of gravitational redshift (otherwise one could introduce an extra constant, or rescale the $t$-coordinate via $\bar t := e^{\alpha(r_0)}t$). Suppose the cable is cut repeatedly, so its end is fixed at $\approx r_\textrm{end}$, which simplifies the calculations. (One could imagine a cutting robot tethered to a static ``service cable''.)
% In some scenarios the winch is geodesic, so $\alpha'(r_0) = 0$, as the forces sum to zero.

\subsection{Kinematics}

Parametrise radial $4$-velocities using the speed $\beta(r)$ relative to a local static observer:
% $u^\mu = \big( Ee^{-2\alpha},\pm\sqrt{E^2e^{-2\alpha}-1}\,e^{-\zeta},0,0 \big)$
\begin{equation}
u^\mu = \gamma\big( e^{-\alpha}, \beta e^{-\zeta}, 0, 0 \big),
\end{equation}
where $\gamma(r) := (1-\beta^2)^{-1/2}$ is the corresponding Lorentz factor, and we allow $\beta < 0$ via $\sgn(\beta) := \sgn(dr/d\tau)$.
% For calculation this is more convenient than $E$, which requires an additional parameter $\pm 1$. Question: Could we use -E as well? E is positive because outside the horizon. However this is potentially confusing, so beta seems better.
The cable's stress-energy is not conserved in general, as mentioned previously, but particle number
\begin{equation}
^{(1)}\nabla_i(nu^i) = 0
\end{equation}
is conserved. Here $n(r)$ is the number density, and the ``$(1)$'' indicates divergence in the radial direction only, which is a sum over $i = 0,1$, in our coordinates. This evaluates to the first-order ODE $(n\beta\gamma)'+(n\beta\gamma)\alpha' = 0$, with solution
\begin{equation}
\label{eqn:nbetagamma}
n\beta\gamma = Ce^{-\alpha}
\end{equation}
for some constant $C$, ignoring trivial cases. In a local static frame, the number flux vector $n\mathbf u$ has components $n\gamma(1,\beta,0,0)$, hence $n\beta\gamma$ is interpreted as the local number flux density in the radial direction, so $C \equiv n\beta\gamma e^\alpha$ is the redshift-corrected value --- that is, the number flux density determined at the winch.
% See Hartle \S22.1 e.483 on number density; MTW e.163

An equation of state gives an additional constraint. Instead of relativistic elasticity, for simplicity we assume Born-rigidity, meaning the expansion tensor is zero (constant proper-frame distance between neighbouring particles) in the radial direction. While Born-rigidity is not possible as an \emph{implicit} property of a physical material, we treat it as a toy model, with discretion.\footnote{Physical intuition is applied here. For instance we derive tension and power, but not the speed of travelling waves. Similarly, our usage of ``tension'' is not necessarily compatible with Lorentz transformation of the stress-energy tensor between frames, but should be self-consistent.}
% physical intuition.. internally self-consistent.. Also, we have ignored the momentum and energy associated with the tension, which may account for itself intrinsically or automatically [\emph{more thought needed}].
% ...due to infinite sound speed... it is permissible if treated as non-causal, for instance with little rockets attached along the material, firing to preserve rigidity.
% \cite{carterquintana1972,karlovinisamuelsson2003}, and citations elsewhere \cite{maugin2013,wernig-pichler2006,beigschmidt2003} Maugin \S15.3.1, Wernig-Pichler \S1,
% cut: Compare Ref.~\citenum{norton1987}  -- this is the logical inconsistency of QM paper

% \theta = \nabla_iu^i = \pm\frac{e^{-\zeta}}{\sqrt{E^2-1}}\big( E'E + \alpha'(E^2-1) \big) \qquad (i=0,1)
Zero divergence $\nabla_iu^i = 0$, $i=0,1$, simplifies to $(\beta\gamma)' + (\beta\gamma)\alpha' = 0$ with solution
\begin{equation}
\label{eqn:K}
\beta\gamma = Ke^{-\alpha}
\end{equation}
for some constant $K$. $\beta\gamma$ is the ``proper speed'' relative to the local static observer, hence $K$ is interpreted as the redshift-corrected proper speed; we treat $K \in (-\infty,\infty)$ as provided. From Equation~\ref{eqn:nbetagamma}, $n = C/K$, and it follows the density $\mu$ is constant. Also,
\begin{equation}
\label{eqn:betagamma}
\gamma = \sqrt{1+K^2e^{-2\alpha}} \qquad\qquad \beta = K/\sqrt{K^2+e^{2\alpha},}
\end{equation}
% previously for V I had divided by K, however this omitted a minus if K<0! Also can have K=0
% Gibbons assumes the tension does not affect rest mass: after his Equation 8 he says ``If $\sigma$ is independent of $T$''
which follows from Equation~\ref{eqn:K}. The cable $4$-velocity is hence
\begin{equation}
\label{eqn:fourvelocity}
u^\mu = e^{-\alpha}(\sqrt{1+K^2e^{-2\alpha}},Ke^{-\zeta},0,0).
\end{equation}
% The Killing energy per mass is
%\begin{equation}
%E = \sqrt{K^2+e^{2\alpha}} \ge e^\alpha
%\end{equation}
%where the minimum (for a given $r$) is obtained by $K = 0$ which is static observers. Inversely, $K = \pm\sqrt{E^2-e^{2\alpha}}$.
Brotas gives similar results for Schwarzschild spacetime.\cite{brotas2006} The $3$-velocity $\vec u$ relative to the local static frame is $(K/\sqrt{K^2+e^{2\alpha}},0,0)$, from the decomposition $\mathbf u = \gamma(1,\vec u)$ in a static frame, with $\beta = \Vert\vec u\Vert$ as expected. While it may seem \emph{a priori} that rigid kinematics are trivial, various authors including Harrison do not correctly treat the frame dependence of length, i.e. ``length-contraction''. Now given the above motion, a cable particle has $4$-acceleration $\mathbf a := \nabla_{\mathbf u}\mathbf u$ with magnitude
\begin{equation}
a = \sqrt{\mathbf a\cdot\mathbf a} = \frac{|\alpha'|e^{-\zeta}}{\sqrt{1+K^2e^{-2\alpha}}}
\end{equation}
directed in the $\sgn(\alpha')$ direction of the $r$-axis. Physically, this acceleration is due to the support of the cable above.

\subsection{Dynamics, energy, and power}

The surplus of energy entering and exiting the system due to the cable's motion gives the power harvested. Consider firstly various energy fluxes measured locally in a static frame at given $r$:
\begin{itemize}
\item mass flux: $\mu\beta_r\gamma_r \equiv \mu Ke^{-\alpha_r}$, locally at $r$%, or $\mu K e^{\alpha(r)-2\alpha(r_1)}$ at $r_1$
\item kinetic energy flux: $\mu\beta_r\gamma_r(\gamma_r-1) \equiv \mu K(\gamma_r-1)e^{-\alpha_r}$, locally
\item mass plus kinetic flux: $\mu\beta_r\gamma_r^2 \equiv \mu K\gamma_r e^{-\alpha_r}$, locally
\end{itemize}
% For mass: the rate is decreased by redshift, also the energy per mass at winch is decreased
% I cut this out: The derivative with respect to $r$ gives the mass flux contributed in the interval $dr$ located at $r$, as measured at $r_1$: $\mu K \alpha'(r)e^{\alpha(r)-2\alpha(r_1)}$
In particular at the winch, the passing cable has mass flux $\mu K$ and kinetic energy flux $\mu K(\sqrt{1+K^2}-1)$, with sum $\mu K\sqrt{1+K^2}$. These are the energy rates entering the system. Now for a given power in a static frame at some $r$, transmitted to a static frame at $r_1$, the received power is redshifted by $e^{\alpha_r-\alpha_1}$ twice. (Imagine transmission via photons, then both the number rate and individual wavelengths are affected.)
% The ``Killing energy per mass'' $E$ for a given observer $\mathbf u$ is $E := -\mathbf u\cdot\partial_t$ which is the energy per mass measured at the winch, assuming $\alpha(r_0) = 0$. Note this comparison between distant frames is only valid because of the time symmetry.
% Incidentally, the measurement of a local static observer, when redshifted, is equivalent to the Killing energy defined from $\boldsymbol\xi$

Subtract the total flux at the start and end of the cable, to determine the overall power as transmitted to the winch frame. This calculation invokes time symmetry. However the incoming kinetic energy must also be provided,\footnote{In practice the cable tension would cause this motion. However in our accounting system, all work done by the cable is harvested, hence the incoming kinetic energy must be considered separately. Also the incoming rest mass is treated as \emph{gratis} or expendable, in order to achieve any profit.} hence at the winch:
% Alternatively, could think of a stockpile of motionless cable, sitting near the winch. Suppose its motion is slow so kinetic energy negligible, but the mass rate passes a fixed point at the same rate as the cable requires. This could be taken as the "true" start of the system.
\begin{equation}
\label{eqn:powerdifference}
\boxed{\mu K\big(1-\sqrt{K^2+e^{2\alpha_\textrm{end}}}\big).}
\end{equation}
% or ..K/\beta_\textrm{end}, assuming $\beta\ne 0$. Or \gamma(r_\textrm{end})e^{\alpha(r_\textrm{end})
% $\mu(\beta\gamma^2)|_{r = r_0} - \mu(\beta\gamma^2e^{2\alpha})|_{r = r_\textrm{end}}$.
The winch end, with redshift $1$, is included implicitly. The term in parentheses is the energy profit per mass, which for a slow cable ending near a Killing horizon ($e^\alpha\approx 0$) approaches $1$, or $100\%$ conversion of $E = mc^2$! If the technology exists to recover the outgoing kinetic energy, the power profit increases to $\mu K(1-e^{\alpha_\textrm{end}})$.

We check using an alternate derivation from forces. The $4$-force on a particle is best defined as $\mathbf f := \nabla_{\mathbf u}\mathbf p$, where $\mathbf p := m\mathbf u$ is the $4$-momentum. It follows $\mathbf f = dm/d\tau\,\mathbf u + m\mathbf a$, but $dm/d\tau = 0$ in our case, a ``pure force''.\cite{tsamparlis2010} In an arbitrary orthonormal frame, $\mathbf f$ has components $\gamma(\dot E,\vec f)$ where $\vec f$ is $3$-force and $\dot E$ is the rate of local energy change in this frame. In a local static frame, $\mathbf f$ has components
% Tsamparlis p.326. This is for a freefalling frame, but I think it is fine for any frame, just consider the static frame instantaneously coinciding with a static frame
\begin{equation}
f^\mu = \gamma\Big( \frac{m\beta^2\alpha'e^{\alpha-\zeta}}{K},\frac{m\alpha'e^{-\zeta}}{\gamma},0,0 \Big),
\end{equation}
% Three-velocity is $(\beta,0,0)$ as already knew, three-force $(m\alpha'e^{-\zeta}/\gamma,0,0)$, inner product: $\beta^2m\alpha'e^{\alpha-\zeta}/K$
which satisfy $\dot E = \vec f\cdot\vec u + \gamma^{-2}dm/d\tau$ as expected.\cite{tsamparlis2010} The $3$-force is unchanged under boosts in the radial direction. Now a coordinate interval $dr$ contains a mass $\mu\gamma e^\zeta dr$ of cable, according to a \emph{static} frame. This is because a static observer measures a proper length $e^\zeta dr$ with its ruler, and so a greater interval $\gamma e^\zeta dr$ of cable fits, according to the cable's length-contracted rulers in this frame.\cite{maclaurin2020MGlength} Replacing $m$ in $\dot E$ with this mass density, the power contributed from this cable element is $\mu\beta\alpha'dr$, which is redshifted twice to the winch. This integrates to $\mu K\sqrt{K^2+e^{2\alpha}}$, which concurs with Equation~\ref{eqn:powerdifference} after evaluating it at the cable ends and subtracting the initial kinetic energy rate. This is the total power gained. The reason for focusing on static frames is that they respect time symmetry, hence measure \emph{sustainable} rates.
% the term can also be written $\mu K\gamma e^\alpha$
% Tsamparlis \S11.2
% on 3-force invariant in radial direction see e.g. Tsamparlis p.331

The $3$-force contributed locally from an interval $dr$ is $\mu\alpha'dr$. After redshifting once to a given $r_1$, then integrating over all cable below $r_1$, the tension at $r_1$ is
\begin{equation}
\boxed{\mu\big(1-e^{\alpha_\textrm{end}-\alpha_1}\big)}
\end{equation}
in the static frame or cable frame, and is independent of $K$. At the winch the tension is $\mu(1-e^{\alpha_\textrm{end}})$. For a cable ending near a Killing horizon, the winch tension is $\approx\mu$, which saturates various energy conditions, and $T/\mu\approx 1$ is the Planck force.
% for $\mu = 1$ (the Planck linear mass density $c^2/G$), the tension is $1$ (the Planck force $c^4/G$).
% purely classical, no h-bar: Planck force, Planck power, Planck linear mass density = Sqrt[hbar c /G] Sqrt[c^3 / hbar G] = c^2/G,  10^(27) kg/m, etc
% force is unitless in geometric units -- Wald e.476. Mass has units [L], hence linear mass density is also dimensionless
% Planck values: https://en.wikipedia.org/wiki/Planck_units#Derived_units

\section{Conclusions and future work}

A cable in curved spacetime illustrates interesting relativistic effects including ``redshift'' of tension. In an idealised scenario, usable energy equivalent to $100\%$ of a cable's rest mass can be extracted; I have extended calculations to a moving cable. The overall Killing energy is conserved. There are many ways this research could be extended. Instead of ignoring backaction, string-like exact solutions could be analysed. Other possible avenues are quantum effects, elasticity, more general motions, application to quasilocal energies, and thermodynamics.
% \cite{anderson2002}

\bibliographystyle{ws-procs975x65}
\bibliography{biblio}

\end{document}